\documentclass[prd,onecolumn,notitlepage,eqsecnum,nofootinbib,floatfix]{revtex4-1}
\usepackage{amssymb}
\usepackage{amsmath}
\usepackage{amsfonts} 
\usepackage{bm}
\DeclareFontFamily{OT1}{rsfs}{} 
\DeclareFontShape{OT1}{rsfs}{m}{n}{<-7> rsfs5 
    <7-10> rsfs7 <10-> rsfs10}{}   
\DeclareMathAlphabet{\scr}{OT1}{rsfs}{m}{n} 
\DeclareSymbolFont{EulerScript}{U}{eus}{m}{n}
\SetSymbolFont{EulerScript}{bold}{U}{eus}{b}{n}
\DeclareSymbolFontAlphabet\scrpt{EulerScript}
\newcommand{\E}{{\cal E}} 
\newcommand{\B}{{\cal B}} 
\newcommand{\F}{{\cal F}} 
\newcommand{\K}{{\cal K}} 
\newcommand{\chid}{\chi^{\scriptstyle \sf d}} 
\newcommand{\Eq}{{\cal E}^{\scriptstyle \sf q}} 
\newcommand{\Bq}{{\cal B}^{\scriptstyle \sf q}} 
\newcommand{\Fd}{{\cal F}^{\scriptstyle \sf d}} 
\newcommand{\Ehatq}{\hat{\cal E}^{\scriptstyle \sf q}} 
\newcommand{\Fo}{{\cal F}^{\scriptstyle \sf o}} 
\newcommand{\Kd}{{\cal K}^{\scriptstyle \sf d}} 
\newcommand{\Bhatq}{\hat{\cal B}^{\scriptstyle \sf q}} 
\newcommand{\Ko}{{\cal K}^{\scriptstyle \sf o}} 
\newcommand{\m}{{\sf m}} 
\newcommand{\eq}{e^{\scriptstyle \sf q}} 
\newcommand{\bq}{b^{\scriptstyle \sf q}} 
\newcommand{\ehatq}{\hat{e}^{\scriptstyle \sf q}}

\newcommand{\bhatq}{\hat{b}^{\scriptstyle \sf q}} 
\newcommand{\kd}{k^{\scriptstyle \sf d}} 
 
\newcommand{\ko}{k^{\scriptstyle \sf o}} 
 
\newcommand{\fd}{f^{\scriptstyle \sf d}} 
\newcommand{\fo}{f^{\scriptstyle \sf o}} 
\newcommand{\cc}[1]{c^{\scriptstyle \sf #1}} 
\newcommand{\gam}[1]{\gamma^{\scriptstyle \sf #1}} 
\newcommand{\ff}[1]{\mathfrak{F}^{\scriptstyle \sf #1}} 
\newcommand{\kk}[1]{\mathfrak{K}^{\scriptstyle \sf #1}} 
\newcommand{\ee}[1]{\mathfrak{E}^{\scriptstyle \sf #1}} 
\newcommand{\bb}[1]{\mathfrak{B}^{\scriptstyle \sf #1}} 
\newcommand{\fff}[1]{\mathfrak{f}^{\scriptstyle \sf #1}} 
\newcommand{\kkk}[1]{\mathfrak{k}^{\scriptstyle \sf #1}} 
\newcommand{\eee}[1]{\mathfrak{e}^{\scriptstyle \sf #1}} 
\newcommand{\bbb}[1]{\mathfrak{b}^{\scriptstyle \sf #1}} 
\newcommand{\ffbar}[1]{\bar{\mathfrak{F}}^{\scriptstyle \sf #1}} 
\newcommand{\kkbar}[1]{\bar{\mathfrak{K}}^{\scriptstyle \sf #1}} 
\newcommand{\eebar}[1]{\bar{\mathfrak{E}}^{\scriptstyle \sf #1}} 
\newcommand{\bbbar}[1]{\bar{\mathfrak{B}}^{\scriptstyle \sf #1}} 
\newcommand{\pd}{p^{\scriptstyle \sf d}} 
\newcommand{\pq}{p^{\scriptstyle \sf q}} 
\newcommand{\po}{p^{\scriptstyle \sf o}} 
\newcommand{\PN}{\mbox{\sc pn}} 
\newcommand{\stf}[1]{{\langle #1 \rangle}} 
\allowdisplaybreaks
\begin{document}
\title{Tidal deformation of a slowly rotating material
  body. External metric}   
\author{Philippe Landry} 
\affiliation{Department of Physics, University of Guelph, Guelph,
  Ontario, N1G 2W1, Canada} 
\author{Eric Poisson} 
\affiliation{Department of Physics, University of Guelph, Guelph,
  Ontario, N1G 2W1, Canada {\ }} 
\affiliation{Canadian Institute for Theoretical Astrophysics,
  University of Toronto, Toronto, Ontario, M5S3H8, Canada} 
\date{April 24, 2015} 
\begin{abstract} 
We construct the external metric of a slowly rotating, tidally
deformed material body in general relativity. The tidal forces acting
on the body are assumed to be weak and to vary slowly with time, and
the metric is obtained as a perturbation of a background metric that
describes the external geometry of an isolated, slowly rotating
body. The tidal environment is generic and characterized by two
symmetric-tracefree tidal moments $\E_{ab}$ and $\B_{ab}$, and the
body is characterized by its mass $M$, its radius $R$, and a
dimensionless angular-momentum vector $\chi^a \ll 1$. The perturbation  
accounts for all couplings between $\chi^a$ and the tidal moments. The
body's gravitational response to the applied tidal field is measured
in part by the familiar gravitational Love numbers $K^{\rm el}_2$ and  
$K^{\rm mag}_2$, but we find that the coupling between the body's
rotation and the tidal environment requires the introduction of four
new quantities, which we designate as rotational-tidal Love numbers.
All these Love numbers are gauge invariant in the usual sense of 
perturbation theory, and all vanish when the body is a black hole.   
\end{abstract} 
\maketitle

\section{Introduction and summary} 
\label{sec:intro} 

In an earlier work (\cite{poisson:15}, hereafter referred to as Paper
0), the geometry of a slowly rotating black hole deformed by tidal
forces was determined under the assumption that the tidal forces are
weak and vary slowly with time. The construction relied on
perturbative techniques, and the metric of the deformed black hole was 
expressed as a perturbation of a background Kerr metric with mass $M$
and dimensionless spin $\chi^a \ll 1$. The perturbation introduces new 
terms to the metric, and these are constructed from tidal moments
$\E_{ab}$ and $\B_{ab}$ that provide, at leading order in the tidal
deformation, a complete characterization of a generic tidal
environment. In particular, the perturbation includes terms that arise
from the coupling between $\chi^a$ and $\E_{ab}$, and the coupling
between $\chi^a$ and $\B_{ab}$; these capture all consequences of the
dragging of inertial frames on the black hole's tidal deformation. The
metric was cast in light-cone coordinates $(v,r,\theta,\phi)$ that
possess a clear geometrical meaning: the advanced-time coordinate $v$
is constant on light cones that converge toward the black hole, the
angular coordinates $(\theta,\phi)$ are constant on the null
generators of each light cone, and the radial coordinate $r$ is an
affine parameter on each generator.   

In this paper (paper I) we replace the black hole by a compact body of
mass $M$, radius $R$, and dimensionless spin $\chi^a \ll 1$, and rely
on the fact that to first order in $\chi^a$, the external geometries
of the unperturbed bodies are identical (differences occur at second
and higher orders). We generalize the perturbative solution of Paper 0
to account for the presence of matter inside $r = R$. The new solution
is no longer required to be regular at $r=2M$ (which marks the
position of the black hole's horizon), and as a consequence we find
that it contains terms that were absent from the black-hole
metric. These describe the body's response to the tidal deformation,
and they come with dimensionless multiplicative constants known as  
{\it gravitational Love numbers}. In the case of a nonrotating body 
($\chi^a = 0$) \cite{damour-nagar:09,binnington-poisson:09,
  landry-poisson:14},  two types of Love numbers appear in the
metric\footnote{A different set of Love numbers, $k^{\rm el}_2$ and
  $k^{\rm mag}_2$, was introduced in
  Ref.~\cite{binnington-poisson:09}. It is related to the set used
  here by $k_2^{\rm el} = (2M/R)^5 K_2^{\rm el}$ and   
  $k_2^{\rm mag} = (2M/R)^5 K_2^{\rm mag}$. The former notation for
  the gravito-magnetic Love number was unfortunate, because Favata 
  \cite{favata:06} has shown that $K_2^{\rm mag}$ scales as $(R/M)^4$   
  instead of $(R/M)^5$.}: a gravito-electric Love number 
$K^{\rm el}_2$ that measures the body's response to the tidal forces
associated with $\E_{ab}$, and a gravito-magnetic Love number 
$K^{\rm mag}_2$ that measures the body's response to those associated
with $\B_{ab}$. With rotation included to first order, we find here
that the description of the external geometry of a tidally deformed
body requires the introduction of four new numbers, which we call
{\it rotational-tidal Love numbers}. The first two, denoted $\ee{q}$
and $\ff{o}$ below, are associated with terms that couple $\chi^a$ to
$\E_{ab}$ in the metric; the remaining two, denoted $\bb{q}$ and
$\kk{o}$, are associated with terms that couple $\chi^a$ to
$\B_{ab}$. The label $\sf q$ indicates that $\ee{q}$ and $\bb{q}$ are
associated with quadrupolar terms in the metric; the label $\sf o$
indicates that $\ff{o}$ and $\kk{o}$ are associated with octupolar
terms. All six Love numbers are gauge-invariant in the usual sense of
perturbation theory: While the form of the solution may change under a 
gauge transformation (understood as an infinitesimal transformation of
the background coordinates), the value of the Love numbers are
preserved.  And our construction guarantees that 
$K^{\rm el}_2 = K^{\rm mag}_2 = \ee{q} = \ff{o} = \bb{q} = \kk{o} = 0$
for a slowly rotating black hole.  

Our considerations in this paper are limited to the exterior geometry
of a slowly rotating, tidally deformed body. In this context, the
new Love numbers $(\ee{q}, \ff{o}, \bb{q}$, $\kk{o})$, like the old
Love numbers $(K^{\rm el}_2, K^{\rm mag}_2)$, must be left
undetermined. The determination of the Love numbers requires the
construction of an internal geometry based on an assumed model for the
body's matter content, and a match of the internal and external
geometries at the matter's boundary; this procedure will be detailed
in Paper II\footnote{Philippe Landry and Eric Poisson, in
  preparation.}. The numerical values adopted by the Love numbers
reflect the details of internal structure, and an external measurement
of the tidal properties of the body can therefore reveal otherwise
inaccessible aspects of the interior.    

This observation has motivated a recent surge of activity in the
development of a relativistic theory of tidal deformation and
dynamics, in the context of the measurement of tidal effects in
gravitational waves emitted by neutron-star binaries 
\cite{flanagan-hinderer:08, hinderer:08, hinderer-etal:10,
  baiotti-etal:10, baiotti-etal:11, vines-flanagan-hinderer:11,
  pannarale-etal:11, lackey-etal:12, damour-nagar-villain:12,
  read-etal:13, vines-flanagan:13, lackey-etal:14, favata:14,
  yagi-yunes:14} and during the capture of solar-mass compact bodies
by supermassive black holes \cite{hughes:01, price-whelan:01,
  martel:04, yunes-etal:10, yunes-etal:11,
  chatziioannou-poisson-yunes:13}.  
The Love number $K^{\rm el}_2$ of a neutron star was implicated, 
along with its moment of inertial $I$ and its rotational quadrupole
moment $Q$, in the remarkable $I$-Love-$Q$ relations
\cite{yagi-yunes:13a, yagi-yunes:13b, doneva-etal:14, maselli-etal:13,
  yagi:14, haskell-etal:14, chakrabarti-etal:14}, and tidal invariants
have been inserted within point-particle actions to account for the
tidal response of an extended body \cite{bini-damour-faye:12,
  chakrabarti-delsate-steinhoff:13a,
  chakrabarti-delsate-steinhoff:13b, dolan-etal:14, bini-damour:14}.    
The geometry of a tidally deformed black hole was constructed beyond
the small rotation limit \cite{yunes-gonzalez:06,
  o'sullivan-hughes:14} and beyond 
perturbation theory \cite{gurlebeck:15}.   

We begin our work in Sec.~\ref{sec:potentials} with the construction
of tidal potentials obtained from $\chi_a$, $\E_{ab}$, and
$\B_{ab}$. These form the fundamental building blocks of the perturbed
metric, which is calculated in Sec.~\ref{sec:metricLC} in the
light-cone coordinates $(v,r,\theta,\phi)$. To illustrate the features
of this metric, it is helpful to introduce Cartesian coordinates 
$x^a = (x,y,z)$ defined in the usual way from the spherical polar 
coordinates $(r,\theta,\phi)$. In this notation, the time-time
component of the metric obtained in Sec.~\ref{sec:metricLC} takes the
form of 
\begin{align} 
g_{vv} &= -1 + \frac{2M}{r} 
- \biggl[ 1 + \cdots 
+ 2 K^{\rm el}_2 \biggl( \frac{2M}{r} \biggr)^5 (1+\cdots) \biggr] 
\E_{ab}\, x^a x^b   
\nonumber \\ & \quad \mbox{} 
+ \frac{1}{2} \biggl( \frac{2M}{r} \biggr)^2 \biggl[ 1+\cdots 
+ 8 \ee{q} \biggl( \frac{2M}{r} \biggr)^3 (1+\cdots) \biggr]
\chi^c \epsilon_{cda} \E^d_{\ b}\, x^a x^b 
\nonumber \\ & \quad \mbox{} 
+ 2M(1 +\cdots) \chi^b \B_{ab}\, x^a
\nonumber \\ & \quad \mbox{} 
- \frac{(2M)^2}{2r^3} \biggl[ 1+\cdots
+ 4 \kk{o} \biggl( \frac{2M}{r} \biggr)^4 (1+\cdots) \biggr] 
\chi_{\langle a} \B_{bc\rangle}\, x^a x^b x^c, 
\end{align} 
in which $\epsilon_{abc}$ is the completely antisymmetric permutation
symbol, all indices are raised and lowered with the Euclidean metric
$\delta_{ab}$, the notation $\langle abc \rangle$ indicates
symmetrization of all indices and removal of all traces, and ellipses
designate relativistic corrections of order $2M/r$ and higher.    

The first two terms in $g_{vv}$ originate from the background metric,
and the following terms proportional to $\E_{ab}\, x^a x^b$ describe a   
quadrupolar tidal deformation of the geometry; the decaying piece
involving $K^{\rm el}_2$ represents the body's response to the applied
tidal field. The next sequence of terms describes another quadrupolar
deformation that arises from the coupling between $\chi^a$
and $\E_{ab}$; the decaying piece involving  $\ee{q}$ represents the
body's response to this coupling. Following this we find a dipolar
deformation that results from the coupling between $\chi^a$ and
$\B_{ab}$. The presence of a dipole indicates that the body is
accelerated in the tidal field, the acceleration being measured by 
$a_a = -M \B_{ab} \chi^b$; this is the Mathisson-Papapetrou
spin force \cite{mathisson:10, papapetrou:51a, papapetrou:51b}
discussed in detail in Paper 0. And finally, the last set of terms
describes an octupolar deformation that also arises from the coupling
between $\chi^a$ and $\B_{ab}$; the decaying piece involving $\kk{o}$
represents the body's response. The time-time component of the metric
features only the Love numbers $K^{\rm el}_2$, $\ee{q}$, and $\kk{o}$;
the remaining Love numbers appear in the remaining components of the
metric.  

The light-cone conditions placed on the metric of
Sec.~\ref{sec:metricLC} only partially determine the
$(v,r,\theta,\phi)$ coordinates, and as a result, the metric features
a number of arbitrary constants that serve to further specify the
choice of gauge. While the coordinate system is still geometrically 
meaningful, the residual gauge freedom implies that the uninteresting
gauge parameters must be determined alongside the physically 
interesting Love numbers, making the task more 
cumbersome than it has to be. (An instance of this additional 
burden can be observed in Ref.~\cite{binnington-poisson:09}.) To
facilitate the determination of the Love numbers in Paper II, in
Sec.~\ref{sec:metricRW} we present another version of the external
metric, this time adopting the standard Boyer-Lindquist
$(t,r,\theta,\phi)$ coordinates for the background metric, and the
familiar Regge-Wheeler gauge for the perturbation.  

As emphasized previously, the complete collection of Love numbers
$(K^{\rm el}_2, K^{\rm mag}_2, \ee{q}, \ff{o}, \bb{q}, \kk{o})$ for a
selected stellar model must be determined by matching an internal
metric to the external metric provided in Sec.~\ref{sec:metricLC} or 
Sec.~\ref{sec:metricRW}. For a given equation of state for cold matter
--- the functions $p(\rho)$ and $\epsilon(\rho)$, in which $\rho$ is
the rest-mass density, $p$ the pressure, and $\epsilon$ the density of
internal (thermodynamic) energy --- the Love numbers can be expressed
as functions of $M/R$, the stellar compactness. While it is known that
$K^{\rm el}_2$ scales as $(R/M)^5$ and $K^{\rm mag}_2$ scales as
$(R/M)^4$ --- modulo relativistic corrections of order $M/R$ and
higher --- our considerations in this paper give us no information
regarding the expected dominant scaling of the remaining Love numbers
with $R/M$. In order to gain some insight into this matter, in
Sec.~\ref{sec:PN} we exploit post-Newtonian methods and attempt to
calculate the Love numbers of a rigidly rotating ball of
incompressible fluid. Based on this calculation, we conclude that
$\ee{q} \sim (R/M)^3$ and $\ff{o} \sim (R/M)^5$ for a generic body
with an arbitrary equation of state. Our results for the remaining
Love numbers are more tentative. Here we conclude that 
$\bb{q} \sim (R/M)^3$ and $\kk{o} \sim (R/M)^4$, with an
admission that one of these relations might be off by one power of
$R/M$; it is thus possible that $\bb{q}$ actually scales 
as $(R/M)^2$, or that $\kk{o}$ actually scales as $(R/M)^3$. These
provisional assignments will be confirmed once a proper matching with
an internal metric is carried out in Paper II. 

As this work was reaching completion we were made aware of an
independent effort \cite{pani-etal:15} to describe the external
geometry of a slowly rotating material body deformed by tidal  
forces. This work generalizes ours in the sense that it constructs the
external metric to second order in the dimensionless spin $\chi^a$. It
is also a restricted version of our own efforts, in the sense that it
allows only for axisymmetric tidal environments with vanishing 
$\B_{ab}$.    

\section{Tidal potentials} 
\label{sec:potentials} 

The construction of tidal potentials is presented in great detail in
Sec.~II of Paper 0 \cite{poisson:15}. Here we summarize the main
results, and introduce the new moments $\hat{\E}_{ab}$ and
$\hat{\B}_{ab}$ that were missed in the earlier work. 

The potentials are obtained by combining $\chi_a$, $\E_{ab}$,
$\B_{ab}$, and $\Omega^a := x^a/r = [\sin\theta\cos\phi,
\sin\theta\sin\phi, \cos\theta]$ in various irreducible ways, with
each potential carrying a specific multipole order $\ell$ and a
specific parity label (even or odd). The coupling of $\chi_a$ and
$\E_{ab}$ produces the odd-parity tensors    
\begin{equation} 
\F_a := \E_{ab} \chi^b, \qquad 
\F_{abc} := \E_{\langle a b} \chi_{c\rangle}, 
\label{F_def} 
\end{equation}  
in which the angular brackets designate the operation of
symmetrization and trace removal, so that $\F_{abc}$ is a
symmetric-tracefree (STF) tensor. It produces also the even-parity
tensor  
\begin{equation} 
\hat{\E}_{ab} := 2 \chi^c \epsilon_{cd(a} \E^d_{\ b)}, 
\label{Ehat_def} 
\end{equation} 
in which $\epsilon_{abc}$ is the antisymmetric permutation symbol. 
The coupling of $\chi_a$ and $\B_{ab}$ produces the even-parity
tensors   
\begin{equation} 
\K_a := \B_{ab} \chi^b, \qquad 
\K_{abc} := \B_{\langle a b} \chi_{c\rangle}.  
\label{K_def} 
\end{equation} 
and the odd-parity tensor 
\begin{equation} 
\hat{\B}_{ab} := 2 \chi^c \epsilon_{cd(a} \B^d_{\ b)}.  
\label{Bhat_def} 
\end{equation} 
The independent components of the STF tensors $\E_{ab}$, $\B_{ab}$,
$\F_a$, $\hat{\E}_{ab}$, $\F_{abc}$, $\K_a$, $\hat{\B}_{ab}$, and
$\K_{abc}$ can be packaged in spherical-harmonic coefficients
$\Eq_\m$, $\Bq_\m$, $\Fd_\m$, $\Ehatq_\m$, $\Fo_\m$, $\Kd_\m$, 
$\Bhatq_\m$, and $\Ko_\m$, respectively. The definitions are
summarized in Table~\ref{tab:coeffs}. 

%\begingroup
%\squeezetable
\begin{table}
\caption{Spherical-harmonic coefficients of STF tensors. 
  The relations between $\Kd_\m$,  $\K_a$, and $\chi \Bq_\m$ are
  identical to those between $\Fd_\m$, $\F_a$, and $\chi
  \Eq_\m$. The relations between $\Bhatq_\m$, $\hat{\B}_{ab}$, and
  $\chi \Bq_\m$ are identical to those between $\Ehatq_\m$,
  $\hat{\E}_{ab}$, and $\chi \Eq_\m$. Finally, the relations between
  $\Ko_\m$, $\K_{abc}$, and $\chi \Bq_\m$ are identical to those
  between $\Fo_\m$, $\F_{abc}$, and $\chi \Eq_\m$.}   
\begin{ruledtabular} 
\begin{tabular}{l} 
$ \chid_0 = \chi_3 = \chi$ \\ 
$ \chid_{1c} = \chi_1 = 0$ \\ 
$ \chid_{1s} = \chi_2 = 0$ \\ 
\\ 
$ \Eq_0 = \frac{1}{2} (\E_{11} + \E_{22}) $ \\ 
$ \Eq_{1c} = \E_{13} $ \\ 
$ \Eq_{1s} = \E_{23} $ \\ 
$ \Eq_{2c} = \frac{1}{2} ( \E_{11} - \E_{22}) $ \\ 
$ \Eq_{2s} = \E_{12} $ \\ 
\\ 
$ \Fd_0 = \F_3 = -2\chi \Eq_0 $ \\ 
$ \Fd_{1c} = \F_1 = \chi \Eq_{1c} $ \\ 
$ \Fd_{1s} = \F_2 = \chi \Eq_{1s} $ \\ 
\\ 
$ \Ehatq_0 = \frac{1}{2} (\hat{\E}_{11} + \hat{\E}_{22}) = 0$ \\ 
$ \Ehatq_{1c} = \hat{\E}_{13} = -\chi \Eq_{1s} $ \\ 
$ \Ehatq_{1s} = \hat{\E}_{23} = \chi \Eq_{1c} $ \\ 
$ \Ehatq_{2c} = \frac{1}{2} ( \hat{\E}_{11} - \hat{\E}_{22}) 
= -2\chi \Eq_{2s} $ \\ 
$ \Ehatq_{2s} = \hat{\E}_{12} = 2 \chi \Eq_{2c} $ \\ 
\\
$ \Fo_0 = \frac{1}{2} (\F_{113} + \F_{223}) 
= \frac{3}{5} \chi \Eq_{0} $ \\ 
$ \Fo_{1c} = \frac{1}{2} (\F_{111} + \F_{122}) 
= -\frac{4}{15} \chi \Eq_{1c} $ \\  
$ \Fo_{1s} = \frac{1}{2} (\F_{112} + \F_{222}) 
= -\frac{4}{15} \chi \Eq_{1s} $ \\  
$ \Fo_{2c} = \frac{1}{2} (\F_{113} - \F_{223}) 
= \frac{1}{3} \chi \Eq_{2c} $ \\  
$ \Fo_{2s} = \F_{123} = \frac{1}{3} \chi \Eq_{2s} $ \\  
$ \Fo_{3c} = \frac{1}{4} (\F_{111} - 3\F_{122}) = 0 $ \\ 
$ \Fo_{2s} = \frac{1}{4} (3\F_{112} - F_{222}) = 0 $
\end{tabular}
\end{ruledtabular} 
\label{tab:coeffs} 
\end{table} 
%\endgroup

%\begingroup
%\squeezetable
\begin{table}
\caption{Spherical-harmonic functions $Y^{\ell\m}$. The functions are
  real, and they are listed for the relevant modes $\ell=1$ (dipole),
  $\ell=2$ (quadrupole), and $\ell=3$ (octupole). The abstract index
  $\m$ describes the dependence of these functions on the angle
  $\phi$; for example $Y^{\ell,2s}$ is proportional to $\sin2\phi$. To
  simplify the expressions we write $C := \cos\theta$ and $S :=
  \sin\theta$.}  
\begin{ruledtabular} 
\begin{tabular}{l} 
$ Y^{1,0} = C $ \\ 
$ Y^{1,1c} = S \cos\phi $ \\ 
$ Y^{1,1s} = S \sin\phi $ \\ 
\\ 
$ Y^{2,0} = 1-3C^2 $ \\ 
$ Y^{2,1c} = 2SC\cos\phi $ \\ 
$ Y^{2,1s} = 2SC\sin\phi $ \\ 
$ Y^{2,2c} = S^2\cos 2\phi $ \\ 
$ Y^{2,2s} = S^2\sin 2\phi $ \\ 
\\
$ Y^{3,0} = C(3-5C^2) $ \\  
$ Y^{3,1c} = \frac{3}{2} S(1-5C^2)\cos\phi $ \\  
$ Y^{3,1s} = \frac{3}{2} S(1-5C^2)\sin\phi $ \\  
$ Y^{3,2c} = 3S^2 C \cos 2\phi $ \\  
$ Y^{3,2s} = 3S^2 C \sin 2\phi $ \\  
$ Y^{3,3c} = S^3 \cos 3\phi $ \\  
$ Y^{3,3s} = S^3 \sin 3\phi $
\end{tabular}
\end{ruledtabular} 
\label{tab:Ylm} 
\end{table} 
%\endgroup

The tidal potentials are decomposed in scalar, vector, and tensor
spherical harmonics, functions of the angular coordinates 
$\theta^A = (\theta,\phi)$. The decomposition involves the scalar
harmonics of Table~\ref{tab:Ylm}, the even-parity harmonics  
\begin{equation} 
Y_A^{\ell\m} := D_A Y^{\ell\m}, \qquad 
Y_{AB}^{\ell\m} := \Bigl[ D_A D_B + \frac{1}{2} \ell(\ell+1)
\Omega_{AB} \Bigr] Y^{\ell\m}, 
\label{even_harm} 
\end{equation} 
and the odd-parity harmonics 
\begin{equation} 
X_A^{\ell\m} := -\epsilon_A^{\ B} D_B Y^{\ell\m}, \qquad 
X_{AB}^{\ell,\m} := -\frac{1}{2} \bigl( \epsilon_A^{\ C} D_B 
+ \epsilon_B^{\ C} D_A \bigr) D_C Y^{\ell\m}. 
\label{odd_harm} 
\end{equation}  
Here $\Omega_{AB} = \mbox{diag}(1,\sin^2\theta)$ is the metric on a
unit two-sphere, and $D_A$ is the covariant-derivative operator
compatible with this metric; $\epsilon_{AB}$ is the Levi-Civita tensor
on the unit two-sphere ($\epsilon_{\theta\phi} = \sin\theta$), and its
index is raised with $\Omega^{AB}$, the matrix inverse to
$\Omega_{AB}$. It should be noted that the tensorial harmonics are
tracefree, in the sense that $\Omega^{AB} Y^{\ell\m}_{AB} = 
\Omega^{AB} X^{\ell\m}_{AB} = 0$. 

The decomposition of the tidal potentials in spherical harmonics is 
described by 
\begin{subequations} 
\label{sph_harm_decomp2}  
\begin{align}
& \chid_A = \sum_\m \chid_\m X^{1\m}_A, \\
& \Fd_A = \sum_\m \Fd_\m X^{1\m}_A, \\
& \Kd = \sum_\m \Kd_\m Y^{1\m}, \qquad 
\Kd_A = \sum_\m \Kd_\m Y^{1\m}_A, \\ 
& \Eq = \sum_\m \Eq_\m Y^{2\m}, \qquad 
\Eq_A = \frac{1}{2} \sum_\m \Eq_\m Y^{2\m}_A, \qquad
\Eq_{AB} = \sum_\m \Eq_\m Y^{2\m}_{AB}, \\ 
& \Ehatq = \sum_\m \Ehatq_\m Y^{2\m}, \qquad 
\Ehatq_A = \frac{1}{2} \sum_\m \Ehatq_\m Y^{2\m}_A, \qquad
\Ehatq_{AB} = \sum_\m \Eq_\m Y^{2\m}_{AB}, \\ 
& \Bq_A = \frac{1}{2} \sum_\m \Bq_\m X^{2\m}_A, \qquad 
\Bq_{AB} = \sum_\m \Bq_\m X^{2\m}_{AB}, \\ 
& \Bhatq_A = \frac{1}{2} \sum_\m \Bhatq_\m X^{2\m}_A, \qquad 
\Bhatq_{AB} = \sum_\m \Bhatq_\m X^{2\m}_{AB}, \\ 
&  \Fo_A = \frac{1}{3} \sum_\m \Fo_\m X^{3\m}_A, \qquad 
\Fo_{AB} = \frac{1}{3} \sum_\m \Fo_\m X^{3\m}_{AB}, \\
& \Ko = \sum_\m \Ko_\m Y^{3\m}, \qquad 
\Ko_A = \frac{1}{3} \sum_\m \Ko_\m Y^{3\m}_A, \qquad 
\Ko_{AB} = \frac{1}{3} \sum_\m \Ko_\m Y^{3\m}_{AB}. 
\end{align} 
\end{subequations} 
It should be noted that tensorial potentials are not defined when
$\ell = 1$. The relations between $\Ehatq_\m$ and $\chi \Eq_\m$
displayed in Table~\ref{tab:coeffs} imply that the tidal potentials
associated with $\hat{\E}_{ab}$ can also be expressed as 
\begin{equation} 
\Ehatq = -\chi \partial_\phi \Eq, \qquad 
\Ehatq_A = -\chi \partial_\phi \Eq_A, \qquad 
\Ehatq_{AB} = -\chi \partial_\phi \Eq_{AB}. 
\end{equation} 
Similarly, we have that 
\begin{equation} 
\Bhatq_A = -\chi \partial_\phi \Bq_A, \qquad 
\Bhatq_{AB} = -\chi \partial_\phi \Bq_{AB}. 
\end{equation} 
In Paper 0 the metric of a slowly rotating, tidally deformed black
hole was presented in terms of the $\phi$-differentiated potentials,
without recognizing that these are in fact associated with the moments
of Eqs.~(\ref{Ehat_def}) and (\ref{Bhat_def}).  
 
\section{External metric of a tidally deformed body --- Light-cone
  gauge}    
\label{sec:metricLC} 

The external metric of an isolated, slowly rotating body of mass $M$
and dimensionless spin $\chi := |\chi^a| \ll 1$ can be expressed as 
\begin{equation} 
ds^2 = -f\, dv^2 + 2\, dvdr + r^2 d\Omega^2 
- 2\frac{2\chi M^2}{r} \sin^2\theta\, dv d\phi, 
\label{background_metric_LC} 
\end{equation} 
where $f := 1-2M/r$ and 
$d\Omega^2 := \Omega_{AB} d\theta^A d\theta^B 
:= d\theta^2 + \sin^2\theta\, d\phi^2$. The metric is displayed in 
coordinates $(v,r,\theta,\phi)$ that are tied to the behavior of
incoming null geodesics that are tangent to converging light cones. It
is easy to show that each surface $v = \mbox{constant}$ is a null
hypersurface, and that its null generators move with constant values
of $\theta$ and $\phi$; $-r$ is an affine parameter on each null
geodesic. 

The metric of a slowly rotating body immersed in a tidal field
produced by remote matter is obtained by perturbing the background 
metric of Eq.~(\ref{background_metric_LC}). The methods to construct  
the perturbation are described in detail in Paper 0 \cite{poisson:15}, 
for the specific case in which the body is a black hole. We adopt
these methods here (with very few details provided), and make the
required changes to account for the presence of matter and the absence
of an event horizon. 

In this section we continue to work in light-cone coordinates, so that
the coordinates $(v,r,\theta,\phi)$ keep their geometrical meaning in
the perturbed spacetime: $v$ continues to be constant on each
converging light cone, $\theta$ and $\phi$ continue to be constant on
each generator, and $-r$ continues to be an affine parameter on each
generator. These requirements imply that 
$g_{vr} = 1$, $g_{rr} = 0 = g_{rA}$, so that $g_{vv}$, $g_{vr}$,
$g_{vA}$, and $g_{AB}$ are the only nonvanishing components of the
metric.   

The perturbed metric is written as 
\begin{subequations} 
\label{metricLC} 
\begin{align} 
g_{vv} &= -f 
- r^2 \eq_1\, \Eq 
+ r^2 \ehatq_1\, \Ehatq 
+ r^2 \kd_1\, \Kd - r^2 \ko_1\, \Ko, \\ 
g_{vr} &= 1, \\ 
g_{vA} &= \frac{2M^2}{r} \chid_A 
- \frac{2}{3} r^3 \bigl( \eq_4\, \Eq_A - \bq_4\, \Bq_A \bigr) 
+ r^3 \bigl( \ehatq_4 \Ehatq_A 
  - \bhatq_4\, \Bhatq_A \bigr)  
- r^3 \bigl( \fd_4\, \Fd_A - \kd_4\, \Kd_A \bigr) 
+ r^3 \bigl( \fo_4\, \Fo_A + \ko_4\, \Ko_A \bigr), \\ 
g_{AB} &= r^2 \Omega_{AB} 
- \frac{1}{3} r^4 \bigl( \eq_7\, \Eq_{AB} - \bq_7\, \Bq_{AB} \bigr) 
+ r^4 \bigl( \ehatq_7\, \Ehatq_{AB} - \bhatq_7\, \Bhatq_{AB} \bigr)
- r^4 \bigl( \fo_7\, \Fo_{AB} - \ko_7\, \Ko_{AB} \bigr), 
\end{align} 
\end{subequations} 
in which $\eq_n$, $\bq_n$ $\ehatq_n$, $\bhatq_n$, $\kd_n$, $\ko_n$,
$\fd_n$, and $\fo_n$ are functions of $r$ that are determined by
solving the vacuum Einstein field equations. The radial functions are
listed in Table~\ref{tab:radialLC}.   

%\begingroup
%\squeezetable
\begin{table}
\caption{Radial functions appearing in the metric of
  Eq.~(\ref{metricLC}), expressed in terms of $x := r/(2M)$, 
  $f := 1-1/x$, and a number of integration constants. All functions
  within square brackets behave as $1 + O(1/x)$ when $x \gg 1$.}   
\begin{ruledtabular} 
\begin{tabular}{l}   
$ \eq_1 = f^2
+ \frac{2}{x^5} \bigl[ -30x^3(x-1)^2\ln f
+ \frac{5}{2}x(2x-1)(1+6x-6x^2) \big] K_2^{\rm el} $ \\ 
$ \eq_4 = f
- \frac{3}{x^5} \bigl[ 20x^4(x-1)\ln f
- \frac{5}{3}x(1+2x+6x^2-12x^3) \bigr] K_2^{\rm el} $ \\ 
$ \eq_7 = 1- \frac{1}{2x^2} 
+ \frac{2}{x^5} \bigl[ -15x^3(2x^2-1)\ln f
+ 5x^2(1-3x-6x^2) \bigr] K_2^{\rm el} $ \\ 
\\
$ \bq_4 = f 
- \frac{3}{x^5} \bigl[ 20x^4(x-1) \ln f
- \frac{5}{3} x(1+2x+6x^2-12x^3) \bigr] K_2^{\rm mag} $ \\ 
$ \bq_7 = 1- \frac{c}{4x^2} 
+ \frac{2}{x^5} \bigl[ -15x^3(2x^2-1) \ln f
+ 5x^2(1-3x-6x^2) \bigr] K_2^{\rm mag} $ \\ 
\\
$ \ehatq_1 = \frac{1}{2x^7} \bigl[ 60x^5(x-1)^2 (\ln f)^2
+ 10x^3(x-1)(1-9x+12x^2) \ln f
- \frac{5}{6}x(1-48x^2+108x^3-72x^4) \bigr] K_2^{\rm el} $ \\ 
$ \qquad \mbox{} 
+ \frac{2}{x^5} \bigl[ -30x^3(x-1)^2 \ln f
+ \frac{5}{2}x(2x-1)(1+6x-6x^2) \bigr] \ee{q} 
+ f^2 \eebar{q} 
- \frac{1}{16 x^4} \gam{q} 
+ \frac{1}{4x^2} - \frac{1}{2x^3} + \frac{1}{8x^4}  $ \\ 
$ \ehatq_4 = -\frac{7}{18x^7} \bigl[
-\frac{360}{7}x^6(x-1) (\ln f)^2
+ \frac{60}{7}x^4(1+10x-12x^2)\ln f
+ \frac{10}{7}x^2(2+9x+24x^2-36x^3) \bigr] K_2^{\rm el} $ \\ 
$ \qquad \mbox{} 
- \frac{2}{x^5} \bigl[ 20x^4(x-1) \ln f 
- \frac{5}{3}x(1+2x+6x^2-12x^3) \bigr] \ee{q} 
+ \frac{2}{3} f \eebar{q} 
+ \frac{1}{24x^4}(2x+1) \gam{q} 
+ \frac{2}{9x^2} - \frac{1}{36x^4} $ \\ 
$ \ehatq_7 = \frac{1}{4 x^7} \bigl[ 20x^5(2x^2-1) (\ln f)^2
- \frac{20}{3} x^4(1+4x)(1-3x) \ln f
+ \frac{10}{9} x^2(1-6x^2+36x^3) \bigr] K_2^{\rm el} $ \\ 
$ \qquad \mbox{} 
+ \frac{2}{3x^5} \bigl[ -15x^3(2x^2-1) \ln f 
+ 5x^2(1-3x-6x^2) \bigr] \ee{q} 
+ \frac{1}{3}(1 - \frac{1}{2x^2}) \eebar{q} 
+ \frac{1}{24 x^3} \gam{q} +\frac{7}{36x^2} $ \\ 
\\
$ \bhatq_4 = -\frac{7}{18x^7} \bigl[ 
-\frac{360}{7}x^6(x-1) (\ln f)^2
+ \frac{60}{7}x^4(1+10x-12x^2) \ln f
+ \frac{10}{7}x^2(2+9x+24x^2-36x^3) \bigr] K_2^{\rm mag} $ \\ 
$ \qquad \mbox{} 
- \frac{2}{x^5} \bigl[ 20x^4(x-1) \ln f 
- \frac{5}{3}x(1+2x+6x^2-12x^3) \bigr] \bb{q} 
+ \frac{2}{3} f\bbbar{q} 
+ \frac{2}{9x^2} - \frac{1}{18x^3} 
- \frac{1}{36x^4} - \frac{c}{72x^4} $ \\
$ \bhatq_7 = \frac{1}{4x^7} \bigl[ 20x^5(2x^2-1) (\ln f)^2
+ \frac{20}{3}x^4(4x+1)(3x-1) \ln f
+ \frac{10}{9}x^2(1-6x^2+36x^3) \bigr] K_2^{\rm mag} $ \\ 
$ \qquad \mbox{}  
+ \frac{2}{3x^5} \bigl[ -15x^3(2x^2-1) \ln f
+ 5x^2(1-3x-6x^2) \bigr] \bb{q} 
+  \frac{1}{3} (1 - \frac{c}{4x^2}) \bbbar{q} 
+ \frac{1}{4x^2} \cc{q} - \frac{1}{36x^3} $ \\
\\
$ \kd_1 = -\frac{2}{5x^7} \bigl[ 15x^3(x-1)(2x-1)(5x-1) \ln f
- \frac{5}{4} x(1+2x-34x^2+144x^3-120x^4) \bigr] K_2^{\rm mag} $ \\ 
$ \qquad \mbox{} 
+ \frac{1}{16 x^4} \cc{d} 
+ \frac{1}{x} - \frac{17}{10x^2} + \frac{4}{5x^3} $ \\
$ \kd_4 = \frac{1}{5x^7} \bigl[ -30x^5(5x-4) \ln f
- \frac{5}{2}x^2(2x-1)(1+6x+30x^2) \bigr] K_2^{\rm mag} $ \\ 
$ \qquad \mbox{} 
- \frac{1}{16x^4} \cc{d} +\frac{1}{2x} - \frac{2}{5x^2} 
- \frac{c}{20x^4} $ \\ 
$ \ko_1 = \frac{1}{x^7} \bigl[
-10x^3(x-1)(1+3x+140x^2-420x^3+280x^4)\ln f
+ \frac{5}{6} x(1+12x+34x^2+244x^3-3640x^4+6720x^5
- 3360x^6) \bigr] K_2^{\rm mag}  $ \\ 
$ \qquad \mbox{} 
+ \frac{2}{x^6} \bigl[ -420x^4(2x-1)(x-1)^2 \ln f
+ 7x^2(1+10x-130x^2+240x^3-120x^4) \bigr] \kk{o} 
+ 2 f^2 (1 - \frac{1}{2x}) x \kkbar{o} 
+ \frac{1}{16x^4} \cc{o} + \frac{1}{2x^2} - \frac{1}{3x^3} $ \\ 
$ \ko_4 = \frac{23}{12 x^7} \bigl[
\frac{60}{23}x^4(5-4x+280x^2-700x^3+420x^4) \ln f
+ \frac{10}{23}x^2(7+32x+46x^2+420x^3-2940x^4
+ 2520x^5) \bigr] K_2^{\rm mag} $ \\ 
$ \qquad \mbox{} 
+ \frac{2}{x^6} \bigl[ 210x^5(3x-2)(x-1) \ln f
+ \frac{7}{2}x^2(1+5x+30x^2-210x^3+180x^4) \bigr] \kk{o} 
- \frac{3}{2} f (1 - \frac{2}{3x}) x \kkbar{o}
+ \frac{5x+1}{32 x^4} \cc{o} 
+ \frac{c}{24x^4} + \frac{1}{3x^2} $ \\ 
$ \ko_7 = -\frac{3}{2x^7} \bigl[ 
-\frac{20}{3}x^4(1+12x-140x^3+140x^4) \ln f
+ \frac{10}{9}x^2(1-2x^2+140x^3+420x^4
- 840x^5) \bigr] K_2^{\rm mag} $ \\ 
$ \qquad \mbox{} 
- \frac{1}{x^6} \bigl[ -84x^4(1-10x^2+10x^3) \ln f
-14x^3(1-10x-30x^2+60x^3) \bigr] \kk{o} 
- (1 - \frac{1}{x} + \frac{1}{10x^3}) x \kkbar{o} 
+ \frac{1}{16x^3} \cc{o} +\frac{1}{3x^2} $ \\
\\
$ \fd_4 = -\frac{1}{5x^7} \bigl[ -30x^5(5x-4) \ln f
- \frac{5}{2} x^2 (2x-1)(1+6x+30x^2) \bigr] K_2^{\rm el} 
+ \frac{1}{x^4} \ffbar{d} + \frac{1}{2x}\gam{d} + \frac{2}{5x^2} $ \\  
$ \fo_4 = -\frac{10}{3x^6} \bigl[ \frac{3}{2}x^3(4x-5) \ln f
- \frac{1}{4}x(7+18x-24x^2) \bigr] K_2^{\rm el} $ \\ 
$ \qquad \mbox{}   
+ \frac{2}{x^6} \bigl[ 210x^5(3x-2)(x-1) \ln f 
+ \frac{7}{2}x^2(1+5x+30x^2-210x^3+180x^4) \bigr] \ff{o} 
+ \frac{3}{2} f (1 - \frac{2}{3x}) x \ffbar{o} 
+ \frac{1}{3x^2} - \frac{5}{12x^3} $ \\  
$ \fo_7 = -\frac{5}{3x^6} \bigl[ -6x^3(2x-1)\ln f
- x(1+12x^2) \bigr] K_2^{\rm el} $ \\ 
$ \qquad \mbox{} 
+ \frac{1}{x^6} \bigl[ -84x^4(1-10x^2+10x^3)\ln f 
- 14x^3(1-10x-30x^2+60x^3) \bigr] \ff{o} 
- (1 - \frac{1}{x} - \frac{1}{10x^3} )x \ffbar{o} 
+ \frac{1}{4x^2} \gam{o} + \frac{1}{6x^3} $
\end{tabular}
\end{ruledtabular} 
\label{tab:radialLC} 
\end{table} 
%\endgroup

The strategy to solve the field equations is as follows. First, we
switch perspectives and consider the metric of Eq.~(\ref{metricLC}) to
be a perturbation of the Schwarzschild solution expressed in
$(v,r,\theta,\phi)$ coordinates --- this is 
Eq.~(\ref{background_metric_LC}) with $\chi$ set equal to zero.  

Second, we  manufacture a first-order perturbation to this new
background metric by introducing the rotational potential $\chid_A$  
and the purely tidal potentials constructed from $\E_{ab}$ and
$\B_{ab}$. With the spherical harmonic decompositions of
Eqs.~(\ref{sph_harm_decomp2}), the linearized field equations 
reduce to a system of coupled differential equations for the radial
functions inserted in front of these potentials. Schematically, these
take the form of  
\begin{equation} 
{\scr L}^j_{\ k} w_1^k = 0, 
\label{pert_eqns_first} 
\end{equation} 
where ${\scr L}^j_{\ k}$ is a second-order differential operator, and
$w_1^j(r)$ is the collection of radial functions that appear in the
first-order metric perturbation. (In practice, the equations for the
rotational, $\ell=1$ perturbation decouple from the equations for the
tidal, $\ell=2$ perturbations, and the equations for $\{ \eq_1, \eq_4,
\eq_7 \}$ decouple from the equations for $\{ \bq_4, \bq_7 \}$.)
Equation (\ref{pert_eqns_first}) is then integrated, and the general
solution is formed from two linearly independent modes, one decaying
as $r$ increases, the other growing. In the case of the rotational
perturbation, the decaying mode produces the angular-momentum term in
Eq.~(\ref{background_metric_LC}), and the growing mode is set to zero,
because it corresponds to an uninteresting coordinate transformation
to a rotating frame. In the case of the tidal perturbation created by
$\E_{ab}$ (respectively $\B_{ab}$), the growing mode represents the
external tidal field, and the decaying mode represents the body's
response to this tidal field, measured by the gravitational Love
number $K_2^{\rm el}$ (respectively $K_2^{\rm mag}$), which appears in
the solution as an integration constant. The solution for $\bq_7$ is
also observed to involve $c$, an additional integration constant that
represents a residual gauge freedom that preserves the geometrical
meaning of the $(v,r,\theta,\phi)$ coordinates.    

Third, the first-order perturbation is used as a seed to construct a
second-order perturbation that accounts for the coupling between
$\chi^a$ and $\E_{ab}$, and the coupling between $\chi^a$ and
$\B_{ab}$. This perturbation ignores terms quadratic in $\chi^a$, and
terms quadratic in $\E_{ab}$ and $\B_{ab}$. The composition of the
$\ell = 1$ and $\ell = 2$ spherical harmonics produces the dipole,
quadrupole, and octupole potentials listed in
Sec.~\ref{sec:potentials}, and insertion of the radial functions gives
rise to the metric of Eq.~(\ref{metricLC}). Substitution of this
metric in the vacuum field equations yields differential equations of
the schematic form  
\begin{equation} 
{\scr L}^j_{\ k} w_2^k = S^j(w_1^k), 
\label{pert_eqns_second} 
\end{equation} 
where $w_2^j$ is the collection of radial functions that appear in the
second-order perturbation, $S^j$ is a set of source terms constructed
from the first-order radial functions, and ${\scr L}^j_{\ k}$ is the
same differential operator as in Eq.~(\ref{pert_eqns_first}). The
general solution to Eq.~(\ref{pert_eqns_second}),  
\begin{equation} 
w_2^j = w_2^j(\text{particular}) + w_2^j(\text{decaying})
+ w_2^j(\text{growing}), 
\end{equation} 
is a linear superposition of a particular solution to the system of
differential equations, a decaying solution to the homogeneous system 
${\cal L}^j_{\ k} w^k_2 = 0$, and a growing solution to the same
homogeneous system\footnote{The growing and decaying solutions can be
  unambiguously identified. The decaying solution decreases with
  increasing $r$, faster than any term in the growing solution, and it
  is singular in the limit $r \to 2M$. The growing solution increases
  with $r$, and is regular in this limit. It should be pointed out
  that the limit $r \to 2M$ is entirely mathematical, since the
  presence of a meterial body inside $r = R > 2M$ implies that 
  $r = 2M$ lies beyond the metric's range of validity. The
  mathematical limit is nevertheless available to identify the growing
  and decaying solutions unambiguously, and the fact that it cannot be
  realized physically is of no concern to this identification.}.     

In Table~\ref{tab:radialLC}, the decaying solutions to the homogeneous 
system are identified with the integration constants $\ee{q}$,
$\ff{o}$, $\bb{q}$, and $\kk{o}$, and the growing solutions are
identified with $\eebar{q}$, $\ffbar{d,o}$, $\bbbar{q}$, and
$\kkbar{o}$. In addition to these, the solutions depend on gauge 
constants $\gam{d,q,o}$ and $\cc{d,q,o}$ that (like $c$) specify  
the residual freedom of the light-cone gauge. It is useful to note
that with $K^{\rm el}_2 = K^{\rm mag}_2 = \ee{q} = \ff{o} = \bb{q} 
= \kk{o}  = \eebar{q} = \ffbar{d,o} = \bbbar{q} = \kkbar{o} = 0$, the
solutions of Table~\ref{tab:radialLC} reduce to those obtained in 
Paper 0 \cite{poisson:15}. 

Many of the integration constants introduced in
Table~\ref{tab:radialLC} do not have a physical meaning, and can be
set equal to zero without loss of generality. We have already put in
this category the gauge constants $c$, $\gam{d,q,o}$, and
$\cc{d,q,o}$, which merely serve to further specify the coordinate
system. We can also single out as unphysical the constants
$\eebar{q}$, $\ffbar{d,o}$, $\bbbar{q}$, and $\kkbar{o}$, which come  
with the growing modes of the radial functions.  

Let us first examine the constant $\ffbar{d}$, which appears in
$\fd_4$. It is easy to see that this term can be removed from $g_{vA}$  
by making the replacement 
\begin{equation} 
\chi^a - 8 M^2 \ffbar{d} \F^a  \to \chi^a 
\end{equation} 
in the metric. This shift of the original spin vector is unobservable,
and its effect on other terms in the metric scale as $\chi^2$ and can
be neglected. Thus, we see that $\ffbar{d}$ can be set equal to zero
without producing a different physical situation. 

Turning next to the terms involving the constants $\eebar{q}$ and
$\bbbar{q}$, we see they also can be removed by performing the shifts  
\begin{equation} 
\E_{ab} - \eebar{q} \hat{\E}_{ab} \to \E_{ab}, \qquad 
\ee{q} - \eebar{q} K_2^{\rm el} \to \ee{q}  
\end{equation} 
and
\begin{equation} 
\B_{ab} - \bbbar{q} \hat{\B}_{ab} \to \B_{ab}, \qquad 
\bb{q} - \bbbar{q} K_2^{\rm mag} \to \bb{q}.  
\end{equation}
These shifts also are unobservable, and their effects on other terms
in the metric can also be neglected. We therefore conclude that
$\eebar{q}$ and $\bbbar{q}$ can be set equal to zero without loss of
generality. 

The case of the constants $\kkbar{o}$ and $\ffbar{o}$ is more subtle,
because the metric of Eq.~(\ref{metricLC}) does not include a tidal
perturbation generated by octupole moments $\E_{abc}$ and $\B_{abc}$
that could be shifted by rotational corrections. But this situation
merely reflects an incompleteness of our description, which can
easily be supplemented with the missing octupole terms (see 
Refs.~\cite{poisson:05, poisson-vlasov:10}). With the octupole
contribution included, we find that the terms involving $\kkbar{o}$
and $\ffbar{o}$ can be eliminated by performing the shifts  
\begin{equation} 
M\E_{abc} + 3 \kkbar{o} \K_{abc} \to M\E_{abc}, \qquad 
\kk{o} - 2 \kkbar{o} K_3^{\rm el} \to \kk{o} 
\end{equation} 
and 
\begin{equation} 
\frac{4}{3} M \B_{abc} + 3 \ffbar{o} \F_{abc}
\to \frac{4}{3} M \B_{abc}, \qquad 
\ff{o} + 2 \ffbar{o} K_3^{\rm mag} \to \ff{o}, 
\end{equation} 
where $K_3^{\rm el}$ and $K_3^{\rm mag}$ are the $\ell=3$
gravitational Love numbers. The factor of $\frac{4}{3}$ in front of
$\B_{abc}$ reflects the definition of the $\ell=3$ tidal potentials
adopted by Poisson and Vlasov \cite{poisson-vlasov:10}; its original
source is the choice of normalization for the tidal moments made in 
Ref.~\cite{zhang:86}. Once again the conclusion is that $\kkbar{o}$ 
and $\ffbar{o}$ can be set equal to zero without altering the physical
situation.      

With the constants $c$, $\gam{d,q,o}$, $\cc{d,q,o}$, $\eebar{q}$,
$\ffbar{d,o}$, $\bbbar{q}$, and $\kkbar{o}$ dismissed as physically
uninteresting, we are left with a metric that depends on the two
original Love numbers $K^{\rm el}_2$ and $K^{\rm mag}_2$, as well as
the remaining four integration constants $\ee{q}$, $\ff{o}$, $\bb{q}$,
and $\kk{o}$. These are attached to the decaying solutions of the
homogeneous perturbation equations, and we interpret them as a new
class of Love numbers associated with the coupling between the body's
rotation and the external tidal field. Fittingly, we may refer to
$\ee{q}$, $\ff{o}$, $\bb{q}$, and $\kk{o}$ as 
{\it rotational-tidal Love numbers}.    

The new Love numbers are thus identified as integration constants in
front of the decaying solutions to the homogeneous perturbation
equations. They are gauge-invariant: While the precise expression of
the decaying solution will be altered by a change of gauge, its
identity as a decaying solution will be preserved, and the numerical
value of $\ee{q}$, $\ff{o}$, $\bb{q}$, and $\kk{o}$ will also be
preserved by the gauge transformation. The gauge-invariant nature of
the Love numbers can also be recognized from a computation of
gauge-invariant quantities. For example, the metric of
Eq.~(\ref{metricLC}) can be used to calculate the Newman-Penrose 
quantity $\psi_0$ in a null tetrad adapted to the principal null
congruence of the slowly rotating background spacetime; as is well
known, $\psi_0$ vanishes in the background spacetime and is therefore
gauge-invariant in the perturbed spacetime. We carried out this
computation but choose not to disclose the results here to avoid
displaying the long expressions. We state nevertheless that $\psi_0$
is observed to depend on $\ee{q}$, $\ff{o}$, $\bb{q}$, and $\kk{o}$,
which each constant multiplying a linearly independent solution of the
Teukolsky equation. Because each term is linearly independent and
therefore individually gauge-invariant, we can conclude that the
rotational-tidal Love numbers are indeed gauge-invariant.   

Why four new Love numbers? As we have seen, the counting of Love
numbers is governed by the coupling between spherical harmonics when
the $\ell = 1$ rotational perturbation is combined with the $\ell = 2$
tidal perturbation in a second-order computation of the complete
perturbation. The coupling produces terms of multipole orders 
$\ell = 1$, $\ell = 2$, and $\ell = 3$, and each term comes with a
number of essential integration constants, which are distinguished
from those associated with the residual gauge freedom. In the case of
$\ell = 2$, the number of essential constants is two for the
even-parity sector, plus two for the odd-parity sector, and we have
seen that two of these can be absorbed into a redefinition of the
tidal quadrupole moments; the two remaining constants are the $\ell=2$
rotational-tidal Love numbers. The conclusion is the same for
$\ell=3$, and we have our total of four new Love numbers. The dipole
case is exceptional, and we have seen that here, the single essential 
constant can be absorbed into a shift of the angular-momentum vector.   

We may generalize the argument and conclude that the coupling between 
the $\ell=1$ rotational perturbation and a tidal perturbation of
multipole order $\ell \geq 3$ will produce six new Love numbers. In
this case the coupling produces terms of multipole order $\ell - 1$,
$\ell$, and $\ell + 1$ in both the even-parity and odd-parity sectors,
with each multipole of each sector coming with two essential
integration constants. One of these can be absorbed into a
redifinition of the associated tidal multipole moment, and we are left
with $3\, \mbox{multipoles} \times 2\, \mbox{sectors} = 6$ new Love 
numbers. 

\section{External metric of a tidally deformed body --- Regge-Wheeler
  gauge}     
\label{sec:metricRW} 

In this section we repeat the calculations carried out in the
preceding section, but express the metric perturbation in the
Regge-Wheeler gauge instead of the light-cone gauge. And instead of
the light-cone coordinates $(v,r,\theta,\phi_{\rm LC})$, we cast the 
background metric in the standard Boyer-Lindquist coordinates
$(t,r,\theta,\phi)$, so that the line element is now expressed as  
\begin{equation} 
ds^2 = -f\, dt^2 + f^{-1}\, dr^2 + r^2 d\Omega^2 
- 2\frac{2\chi M^2}{r} \sin^2\theta\, dt d\phi, 
\label{background_metric_BL} 
\end{equation} 
where $f := 1-2M/r$ and 
$d\Omega^2 := \Omega_{AB} d\theta^A d\theta^B 
:= d\theta^2 + \sin^2\theta\, d\phi^2$. It is important to note that
the coordinate $\phi$ adopted here is distinct from the 
$\phi_{\rm LC}$ featured in Sec.~\ref{sec:metricLC} (and 
denoted $\phi$ there); we have the relations $dv = dt + f^{-1}\, dr$
and $d\phi_{\rm LC} = d\phi + (2\chi M^2/r^3 f)\, dr$ between the
coordinate systems.     

%\begingroup
%\squeezetable
\begin{table}
\caption{Radial functions appearing in the metric of
  Eq.~(\ref{metricRW}), expressed in terms of $x := r/(2M)$, 
  $f := 1-1/x$, and a number of integration constants. All functions
  within square brackets behave as $1 + O(1/x)$ when $x \gg 1$.}   
\begin{ruledtabular} 
\begin{tabular}{l}   
$ \eq_{tt} = f^2 + \frac{2}{x^5} \bigl[ -30x^3(x-1)^2 \ln f 
- \frac{5}{2} x(2x-1)(6x^2-6x-1) \bigr] K_2^{\rm el} $ \\ 
$ \eq_{rr} = f^{-2} \eq_{tt} $ \\ 
$ \eq = 1 - \frac{1}{2x^2}  
+ \frac{2}{x^5} \bigl[ -15x^3(2x^2-1) \ln f 
- 5x^2(6x^2+3x-1) \bigr] K_2^{\rm el} $ \\ 
\\ 
$ \bq_t = f - \frac{3}{x^5} \bigl[ 20x^4(x-1) \ln f 
+ \frac{5}{3} x(12x^3-6x^2-2x-1) \bigr] K_2^{\rm mag} $ \\ 
\\ 
$ \ehatq_{tt} = \frac{2}{x^5} \bigl[ -30x^3(x-1)^2 \ln f 
- \frac{5}{2} x(2x-1)(6x^2-6x-1) \bigr] (\ee{q} - \frac{1}{120}) 
+ f^2 \eebar{q} $ \\ 
$ \ehatq_{tr} = -\frac{1}{3x^7} \bigl[ 15x^4(3x-1)\ln f 
+ \frac{5}{4} \frac{x^2}{x-1}(36x^3-30x^2-1) \bigr] K_2^{\rm el} 
+ \frac{1}{4x^2}  - \frac{1}{12x^3} $ \\ 
$ \ehatq_{rr} = f^{-2} \ehatq_{tt} $ \\ 
$ \ehatq = \frac{2}{x^5} \bigl[ -15x^3(2x^2-1) \ln f 
-5x^2(6x^2+3x-1) \bigr] (\ee{q} - \frac{1}{120}) 
+ \frac{1}{2x^2}(2x^2-1) \eebar{q} $ \\ 
\\ 
$ \bhatq_t = -\frac{2}{x^5} \bigl[ 20x^4(x-1) \ln f 
+ \frac{5}{3} x(12x^3-6x^2-2x-1) \bigr] (\bb{q} - \frac{1}{120})  
+ \frac{2}{3} f \bbbar{q} $ \\ 
$ \bhatq_r = \frac{3}{4x^7} \bigl[ -\frac{20}{3} \frac{x^5(2x-1)}{x-1} \ln f 
- \frac{10}{9} \frac{x^2(12x^3+x+1)}{x-1} \bigr] K_2^{\rm mag} 
+ \frac{1}{12} \frac{2x-1}{x^2(x-1)} $ \\ 
\\ 
$ \kd_{tt} = -\frac{2}{5x^7} \bigl[ \frac{15}{2}(20x-9)(x-1)x^4 \ln f 
+ \frac{5}{8} \frac{x}{x-1}(240x^5-468x^4+242x^3-16x^2-x+2) 
\bigr] K_2^{\rm mag} 
+ \frac{1}{x^3(x-1)} \cc{d}  
+ \frac{1}{20} \frac{20x^2-49x+38}{x^2(x-1)} $ \\ 
$ \kd_{rr} = \frac{6}{5x^7} \bigl[ \frac{15}{2} \frac{x^6}{x-1} \ln f 
+ \frac{5}{8} \frac{x^3}{(x-1)^3} (12x^4-18x^3+4x^2+x+2) 
\bigr] K_2^{\rm mag} 
+ \frac{3}{x(x-1)^3} \cc{d} 
- \frac{3}{20} \frac{(x+2)(x-4)}{x(x-1)^3} $ \\ 
$ \ko_{tt} = \frac{1}{x^7} \bigl[ -10x^4(x-1)(280x^3-420x^2+140x+3)
\ln f  - 2800x^7 + 5600x^6 - \frac{9100}{3} x^5 + \frac{610}{3}x^4
+ \frac{115}{3} x^3 +5x^2 - \frac{5}{6}x -\frac{5}{6} 
\bigr] K_2^{\rm mag} 
$ \\ $ \qquad \mbox{}
+ \frac{2}{x^6} \bigl[ -420x^4(2x-1)(x-1)^2 \ln f 
- 7x^2(120x^4-240x^3+130x^2-10x-1) \bigr] \kk{o} 
+ f^2 (2x-1) \kkbar{o} 
+ \frac{1}{2x^2} - \frac{1}{2x^3} $ \\ 
$ \ko_{rr} = \frac{7}{x^7} \bigl[ -\frac{10}{7} \frac{x^6}{x-1} 
(280x^3-420x^2+140x+1) \ln f 
-\frac{5}{42} \frac{x^2}{(x-1)^2}
(3360x^7-6720x^6+3640x^5-268x^4-34x^3-2x^2+3x-5) \bigr] K_2^{\rm mag}
$ \\ $ \qquad \mbox{}
+ \frac{2}{x^6} \bigl[ -420x^6(2x-1) \ln f 
- 7 \frac{x^4}{(x-1)^2} (120x^4-240x^3+130x^2-10x-1) \bigr] \kk{o} 
+ (2x-1) \kkbar{o} + \frac{1}{6 x(x-1)} $ \\  
$ \ko = -\frac{2}{x^7} \bigl[ -10x^5(140x^3-140x^2+13) \ln f 
- \frac{5}{6} x(1680x^6-840x^5-280x^4+16x^3 -6x^2-4x-1) 
\bigr] K_2^{\rm mag} 
$ \\ $ \qquad \mbox{}
- \frac{2}{x^6} \bigl[ -84x^4(10x^3-10x^2+1) \ln f 
- 14x^3(60x^3-30x^2-10x+1) \bigr] \kk{o} 
- \frac{1}{5x^2}(10x^3-10x^2+1) \kkbar{o} 
+ \frac{1}{3x^2} $ \\ 
\\ 
$ \fd_t = -\frac{1}{28x^8} \bigl[ -84x^4(10x^3-10x^2+1) \ln f 
-14x^3(60x^3-30x^2-10x+1) \bigr] K_2^{\rm el} 
+ \frac{1}{x^4} \ffbar{d} + \frac{1}{2x} \gam{d} + \frac{1}{2x^2} $ \\ 
$ \fo_t = -\frac{10}{3x^6} \bigl[ -\frac{3}{2} x^2(5x-2)\ln f 
- \frac{3}{4} x(10x+1) \bigr] K_2^{\rm el} 
$ \\ $ \qquad \mbox{}
+ \frac{2}{x^6} \bigl[ 210x^5(3x-2)(x-1) \ln f 
+ \frac{7}{2}x^2(180x^4-210x^3+30x^2+5x+1) \bigr] \ff{o} 
+ \frac{1}{2} f (3x-2) \ffbar{o} - \frac{5}{12x^3} + \frac{1}{6x^4} $
\end{tabular}
\end{ruledtabular} 
\label{tab:radialRW} 
\end{table} 
%\endgroup
 
The Regge-Wheeler gauge conditions are formulated at the level of each
$\ell\m$ mode of the metric perturbation. For the even-parity sector
generated by $\E_{ab}$, $\hat{\E}_{ab}$, $\K_{a}$, and $\K_{abc}$, we
decompose the metric perturbation $p_{\alpha\beta}$ as 
\begin{equation} 
p_{ab} = \sum_{\ell\m} h_{ab}^{\ell\m} Y^{\ell\m}, \qquad 
p_{aB} = \sum_{\ell\m}  j_a^{\ell\m} Y^{\ell\m}_A, \qquad 
p_{AB} = r^2 \sum_{\ell\m} \bigl( K^{\ell\m} \Omega_{AB} Y^{\ell\m} 
+ G^{\ell\m} Y^{\ell\m}_{AB} \bigr), 
\end{equation} 
where $x^a = (t,r)$ and $\theta^A = (\theta,\phi)$, and we impose
$j^{\ell\m}_a = 0 = G^{\ell\m}$ for $\ell = 2$ and $\ell = 3$. For
$\ell=1$, the tensorial harmonics $Y^{\ell\m}_{AB}$ and the associated
perturbation variables $G^{\ell\m}$ are not defined, and we adopt 
instead $K^{1,\m} = 0$ as a gauge condition, along with 
$j^{1,\m}_a = 0$. For the odd-parity sector generated by $\B_{ab}$,
$\hat{\B}_{ab}$, $\F_{a}$, and $\F_{abc}$, we decompose the metric
perturbation as   
\begin{equation} 
p_{ab} = 0, \qquad 
p_{aB} = \sum_{\ell\m} h_a^{\ell\m} X^{\ell\m}_A, \qquad 
p_{AB} = \sum_{\ell\m} h_2^{\ell\m} X^{\ell\m}_{AB} 
\end{equation} 
and impose $h_2^{\ell\m} = 0$ for $\ell = 2$ and $\ell = 3$. For
$\ell=1$, $X^{\ell\m}_{AB}$ and $h_2^{\ell\m}$ are not defined, and we
adopt instead $h_r^{1,\m} = 0$ as a gauge condition; this choice is
motivated by the fact that $h_r^{3,m}$ is found to vanish by virtue of
the field equations.  

The strategy to integrate the vacuum Einstein field equations is
readily adapted from Sec.~\ref{sec:metricLC}, and we obtain a
perturbed metric of the form 
\begin{subequations} 
\label{metricRW} 
\begin{align} 
g_{tt} &= -f 
- r^2 \eq_{tt}\, \Eq 
+ r^2 \ehatq_{tt}\, \Ehatq 
+ r^2 \kd_{tt}\, \Kd - r^2 \ko_{tt}\, \Ko, \\ 
g_{tr} &= r^2 \ehatq_{tr}\, \Ehatq, \\ 
g_{rr} &= f^{-1} - r^2 \eq_{rr}\, \Eq 
+ r^2 \ehatq_{rr}\, \Ehatq 
+ r^2 \kd_{rr}\, \Kd - r^2 \ko_{rr}\, \Ko, \\ 
g_{tA} &= \frac{2M^2}{r} \chid_A 
+ \frac{2}{3} r^3 \bq_t\, \Bq_A 
- r^3 \bhatq_t\, \Bhatq_A 
- r^3 \fd_t\, \Fd_A 
+ r^3 \fo_t\, \Fo_A, \\ 
g_{rA} &= -r^3 \bhatq_r\, \Bhatq_A, \\ 
g_{AB} &= r^2 \Omega_{AB} 
- r^4 \eq\, \Eq_{AB} 
+ r^4 \ehatq\, \Omega_{AB} \Ehatq 
+ r^4 \ko\, \Omega_{AB} \Ko, 
\end{align} 
\end{subequations} 
with the radial functions listed in Table~\ref{tab:radialRW}.   

The radial functions depend on a number of integration constants, but
many do not have a physical meaning. As was discussed in
Sec.~\ref{sec:metricLC}, the terms involving $\eebar{q}$,
$\ffbar{d,o}$, $\bbbar{q}$, and $\kkbar{o}$ can be 
eliminated by redefining $\chi^a$, $\E_{ab}$, $\B_{ab}$, $\E_{abc}$,
and $\B_{abc}$; these constants can therefore be set equal to zero
without loss of generality. The constants $\cc{d}$ and $\gam{d}$
specify the residual gauge freedom associated with the $\ell = 1$ 
even-parity and odd-parity modes, respectively. (There is no residual
gauge freedom in the $\ell \geq 2$ modes; in this case the
Regge-Wheeler conditions completely specify the coordinate system.)    

The remaining integration constants,$\ee{q}$, $\ff{o}$, $\bb{q}$, and
$\kk{o}$ can again be interpreted as a new class of rotational-tidal 
Love numbers. Computation of $\psi_0$ from the metric of
Eq.~(\ref{metricRW}) and comparison with the results obtained (but 
not displayed) in Sec.~\ref{sec:metricLC} confirms that the Love
numbers introduced here are the same as those introduced
in Sec.~\ref{sec:metricLC}. This consistency check explains the
curious shifts $\ee{q} \to \ee{q} - \frac{1}{120}$, $\bb{q} \to \bb{q}
- \frac{1}{120}$ observed in Table~\ref{tab:radialRW}; without these
the metric of Eq.~(\ref{metricRW}) would not be directly obtained from
the metric of Eq.~(\ref{metricLC}) after a coordinate transformation
followed by a gauge transformation.   

\section{Slowly rotating, tidally deformed body in post-Newtonian
  theory}  
\label{sec:PN} 

In this section we develop a post-Newtonian description of the
geometry of a slowly rotating, tidally deformed material body. We
perform a leading-order calculation that accounts for the coupling
between $\chi^a$ and $\E_{ab}$; the coupling between $\chi^a$ and
$\B_{ab}$ occurs at a higher post-Newtonian order and is neglected
here. We adopt the simplest model for the slowly rotating body: an 
incompressible fluid with uniform density $\rho_0$, rotating rigidly  
with a constant angular velocity $\omega^a$. Our main objective in
this section is to obtain useful information regarding the scaling of 
$\ee{q}$, $\ff{o}$, $\bb{q}$, and $\kk{o}$ with the body's radius $R$.   

\subsection{Newtonian structure} 

The unperturbed body has a density function $\rho = \rho_0
\Theta(R-r)$, with $R$ denoting the body's radius, an internal 
potential $U_{\rm in} = (GM/2R)(3-r^2/R^2)$, an external potential
$U_{\rm out} = GM/r$, and its pressure profile is given by 
$p = (3GM^2/8\pi R^4)(1-r^2/R^2)$. The body's mass is related to its
density by $M = \frac{4}{3} \pi \rho_0 R^3$. 

The tidal deformation produces perturbations $\delta \rho$, 
$\delta U_{\rm in}$, $\delta U_{\rm out}$, and $\delta p$ to these 
quantities, as well as a perturbation $\delta R$ to the position of
the boundary. With a tidal environment characterized by a quadrupole
moment $\E_{ab}$, we have that $\delta U_{\rm out}$ can be expressed
quite generally as 
\begin{equation} 
\delta U_{\rm out} = -\frac{1}{2} \bigl[ 1 + 2k_2 (R/r)^5 \bigr] 
\E_{ab} x^a x^b,
\label{deltaUout} 
\end{equation} 
in terms of a gravitational Love number $k_2$; the first term (which
grows as $r^2$) represents the external tidal field, while the second
term (decaying as $r^{-3}$) represents the body's response to the
tidal forces. With this information, $\delta R$ is deduced from the
requirement that the boundary must be an equipotential surface;
we obtain 
\begin{equation} 
\delta R = -\frac{R^4}{2GM}  (1+2k_2) 
\E_{ab} \Omega^a \Omega^b. 
\label{deltaR} 
\end{equation} 
The perturbation of the internal potential must have a pure
quadrupolar form proportional to $\E_{ab} x^a x^b$, and the constant
of proportionality is obtained by continuity of $\delta U_{\rm in}$
and $\delta U_{\rm out}$ at $r=R$; we find
\begin{equation} 
\delta U_{\rm in} = -\frac{1}{2} (1 + 2k_2) 
\E_{ab} x^a x^b. 
\label{deltaUin} 
\end{equation} 
The density perturbation is obtained from the new assignment 
$\rho + \delta \rho = \rho_0 \Theta(R + \delta R - r)$. Expansion of
the step function to first order in $\delta R$ and use of
Eq.~(\ref{deltaR}) produces 
\begin{equation} 
\delta \rho = -\frac{3R}{8\pi G} (1+2k_2) \E_{ab} \Omega^a \Omega^b\, 
\delta(r-R). 
\label{deltarho} 
\end{equation} 
To obtain the pressure perturbation we appeal to the statement of 
hydrostatic equilibrium, $\partial_a \delta p = \rho \partial_a 
\delta U_{\rm in} + \delta \rho \partial_a U_{\rm in}$. The angular
components of this equation integrate to $\delta p = \rho_0 
\delta U_{\rm in}$, and we get
\begin{equation} 
\delta p = -\frac{3M}{8\pi R^3} (1+2k_2) \E_{ab} x^a x^b. 
\label{deltap}
\end{equation} 
It can be verified that $p + \delta p$ properly vanishes when  
$r = R + \delta R$. The results for $\delta R$, $\delta \rho$, and
$\delta p$ can alternatively be derived on the basis of the
Lagrangian displacement vector 
\begin{equation} 
\xi_a = -\frac{R^3}{2GM} (1+2k_2) \E_{ab} x^b. 
\label{ladisvec} 
\end{equation} 
Finally, $k_2$ can be determined from the
discontinuity in $\partial_r \delta U$ at $r=R$ implied by Poisson's
equation $\nabla^2 \delta U = -4\pi G \delta \rho$ and the
$\delta$-function of Eq.~(\ref{deltarho}). A short calculation reveals 
that the discontinuity must be given by $[\partial_r \delta U] 
= \frac{3}{2} (1+2k_2) R \E_{ab}  \Omega^a \Omega^b$, 
where $[\partial_r \delta U] := \partial_r U_{\rm out} 
- \partial_r U_{\rm in}$ evaluated at $r=R$. Substituting our previous
results for $\delta U_{\rm in}$ and $\delta U_{\rm out}$, we arrive at 
\begin{equation} 
k_2 = \frac{3}{4}, \qquad 1 + 2k_2 = \frac{5}{2}. 
\label{k2} 
\end{equation} 
This is the well-known value of the gravitational Love number for an 
incompressible fluid. 

The body is taken to be rotating rigidly with a uniform angular
velocity $\omega^a$. The rotation is assumed to be slow, and we work
consistently to first order in $\omega^a$. The unperturbed velocity
field inside the body is $v_a = \epsilon_{abc} \omega^b x^c$, the
spin angular momentum of the unperturbed body is given by 
$S^a = \frac{2}{5} M R^2 \omega^a$, and its dimensionless version is
$\chi^a = (c/GM^2) S^a$, or  
\begin{equation} 
\chi^a = \frac{2}{5} \frac{cR^2}{GM} \omega^a. 
\label{chi_vs_omega}
\end{equation} 
In terms of this the velocity field becomes 
\begin{equation} 
v_a = \frac{5}{2} \frac{GM}{cR^2} \epsilon_{abc} \chi^b x^c. 
\label{v_vs_chi} 
\end{equation} 
The tidal deformation creates a perturbation $\delta v_a =
v^b \partial_b \xi_a - \xi^b \partial_b v_a$ of the velocity field, 
which can be calculated from the Lagrangian displacement 
vector of Eq.~(\ref{ladisvec}). We obtain 
\begin{equation} 
\delta v_a = \frac{25}{8} \frac{R}{c} \hat{\E}_{ab} x^b. 
\label{delta_va} 
\end{equation}   
The tidal perturbation changes the body's angular momentum by 
$\delta S_a = \epsilon_{abc} \int x^b \delta j^c\, d^3x$, where
$\delta j_a = \delta\rho\, v_a + \rho \delta v_a$. A short calculation
reveals that the shift in the dimensionless spin is given by    
\begin{equation} 
\delta \chi^a = \frac{5}{4} \frac{R^3}{GM} \E^a_{\ b} \chi^b 
= \frac{5}{4} \frac{R^3}{GM} \F^a.  
\label{deltachi} 
\end{equation} 

\subsection{Vector potential} 

Our post-Newtonian calculation is based on the metric 
\begin{equation} 
g_{00} = -1 + \frac{2}{c^2} U + O(c^{-4}), \qquad 
g_{0a} = -\frac{4}{c^3} U_a + O(c^{-5}), \qquad 
g_{ab} = \delta_{ab} \biggl( 1 + \frac{2}{c^2} U \biggr) 
+ O(c^{-4}),  
\label{PNmetric} 
\end{equation} 
which involves a vector potential $U_a$ in addition to the Newtonian
potential $U$ encountered previously. The coupling between $\chi^a$
and $\E_{ab}$ is captured by the vector potential, which satisfies the
field equation  
\begin{equation} 
\nabla^2 U_a = -4\pi G \rho v_a. 
\label{Poisson_Ua} 
\end{equation} 
An examination of the post-Newtonian equations (see, for example,
Sec.~8.1 of Ref.~\cite{poisson-will:14}) reveals that the coupling
does not appear in $g_{00}$ at first post-Newtonian ($1\PN$) order,
and we therefore omit the $1\PN$ terms at order $c^{-4}$. Similarly,
consideration of $g_{ab}$ at $2\PN$ order reveals that the coupling
does not appear within the omitted $c^{-4}$ terms. The unperturbed
vector potential is     
\begin{equation} 
U_a^{\rm out} = \frac{(GM)^2}{2c r^3} \epsilon_{abc} \chi^b x^c
\end{equation} 
outside the body, and 
\begin{equation} 
U_a^{\rm in} = \frac{(GM)^2}{4c R^3} (5-3r^2/R^2) 
\epsilon_{abc} \chi^b x^c
\end{equation} 
inside the body. The additional factor of $c^{-1}$ results from
expressing the angular-momentum vector in terms of its dimensionless
version of Eq.~(\ref{chi_vs_omega}); this promotes $g_{0a}$ to a
quantity of $1.5\PN$ order. 

The perturbation of the vector potential is sourced by the sum of 
$\delta_1 j_a := \delta\rho\, v_a$ and 
$\delta_2 j_a = \rho \delta v_a$, and correspondingly it is expressed
as the sum of  
\begin{equation} 
\delta_1 U_a(\bm{x}) = G \int \frac{\delta_1 j_a(\bm{x'})}
  {|\bm{x}-\bm{x'}|}\, d^3x', \qquad   
\delta_2 U_a(\bm{x}) = G \int \frac{\delta_2 j_a(\bm{x'})}
  {|\bm{x}-\bm{x'}|}\, d^3x'.   
\label{deltaUa1} 
\end{equation} 
To evaluate this we rely on the addition theorem 
\begin{equation} 
\frac{1}{|\bm{x}-\bm{x'}|} = \sum_{\ell m} \frac{4\pi}{2\ell + 1} 
\frac{r_<^\ell}{r_>^{\ell+1}} Y^*_{\ell m}(\theta',\phi')
Y_{\ell m}(\theta,\phi), 
\label{addition} 
\end{equation} 
the identity [Eq.~(1.171) of Ref.~\cite{poisson-will:14}; 
$L := a_1 a_2 \cdots a_\ell$ is a multi-index that includes a number
$\ell$ of individual indices, $\Omega^L := \Omega^{a_1} \Omega^{a_2}
\cdots \Omega^{a_\ell}$, and the angular brackets indicate that all
traces are to be removed from $\Omega^L$]    
\begin{equation} 
\sum_m Y_{\ell m}(\theta,\phi) \int Y^*_{\ell m}(\theta',\phi') 
\Omega^{\prime \stf{L'}}\, d\Omega' 
= \delta_{\ell,\ell'} \Omega^\stf{L}, 
\label{identity} 
\end{equation} 
and a decomposition of 
\begin{equation}  
Z_a := \epsilon_{abc} \chi^b \E_{de} \Omega^c \Omega^d \Omega^e   
\end{equation} 
into irreducible components. This is accomplished with the identity 
$\Omega^c \Omega^d \Omega^e = \Omega^\stf{cde} 
+ \frac{1}{5}( \delta^{cd} \Omega^e + \delta^{ce} \Omega^d 
+ \delta^{de} \Omega^c )$, which gives rise to 
$Z_a = \frac{2}{5} Z^1_a + Z^3_a$, where 
\begin{equation}  
Z^1_a := \epsilon_{abc} \chi^b \E^c_{\ d} \Omega^d, \qquad 
Z^3_a := \epsilon_{abc} \chi^b \E_{de} \Omega^\stf{cde}. 
\end{equation} 
These quantities can be expressed in terms of the tidal potentials
\cite{poisson:15} 
\begin{subequations} 
\begin{align} 
\Fd_a &:= \frac{1}{c^2} \epsilon_{abc} \Omega^b \F^c, \qquad   
\Fo_a := \frac{1}{c^2} \epsilon_{abc} \Omega^b 
\F^c_{\ de} \Omega^d \Omega^e, \\ 
\Ehatq &:= \frac{1}{c^2} \hat{\E}_{ab} \Omega^a \Omega^b, \qquad 
\Ehatq_a := \frac{1}{c^2} \left( \delta_a^{\ b} - \Omega_a \Omega^b
\right) \hat{\E}_{bc} \Omega^c,
\end{align} 
\end{subequations} 
which are Cartesian versions of the angular potentials introduced in
Sec.~\ref{sec:potentials}; the relation is given, for example, by
$\Fo_A = \Fo_a \Omega^a_A$, where 
$\Omega^a_A := \partial \Omega^a/\partial \theta^A$. We have 
\begin{equation}
c^{-2} Z^1_a = \frac{1}{2} \bigl( \Fd_a + \Ehatq_a 
+ \Omega_a \Ehatq \bigr), \qquad 
c^{-2} Z^3_a = \frac{1}{15} \bigl( 2 \Ehatq_a  
- 3 \Omega_a \Ehatq \bigr) - \Fo_a. 
\label{Z_vs_potentials} 
\end{equation} 

The computation of $\delta_1 U_a$ and $\delta_2 U_a$ proceeds by
inserting 
\begin{equation} 
\delta_1 j_a = -\frac{75}{32\pi}\frac{M}{c} Z_a\, \delta(r-R), \qquad 
\delta_2 j_a = \frac{75}{32\pi} \frac{M}{cR^2} \hat{\E}_{ab} x^b\,
\Theta(R-r),  
\end{equation} 
as well as the  addition theorem of Eq.~(\ref{addition}), within
Eq.~(\ref{deltaUa1}). The angular integrals are evaluated with the
help of Eq.~(\ref{identity}), after involving the decomposition of
$Z_a$ into $Z_a^1$ and $Z_a^3$. The final expressions,  
\begin{equation} 
\delta_1 U^{\rm out}_a = -\frac{75}{8} GMc \biggl[ 
\frac{1}{15} \frac{R^3}{r^2} \biggl( \Fd_a + \Ehatq_a 
+ \Omega_a \Ehatq \biggr) 
+ \frac{1}{7} \frac{R^5}{r^4} \biggl( \frac{2}{15} \Ehatq_a 
- \frac{1}{5} \Omega_a \Ehatq - \Fo_a \biggr) \biggr]  
\label{delta1Ua2}  
\end{equation} 
and 
\begin{equation} 
\delta_2 U^{\rm out}_a = \frac{75}{8} GMc \biggl[ 
\frac{1}{15} \frac{R^3}{r^2} \biggl( \Ehatq_a 
+ \Omega_a \Ehatq \biggr) \biggr],   
\label{delta2Ua2}  
\end{equation} 
are obtained after making use of Eq.~(\ref{Z_vs_potentials}). The
complete perturbation is
\begin{equation} 
\delta U^{\rm out}_a = -\frac{75}{8} GMc \biggl[ 
\frac{1}{15} \frac{R^3}{r^2} \Fd_a 
+ \frac{1}{105} \frac{R^5}{r^4} \bigl( 2 \Ehatq_a 
- 3 \Omega_a \Ehatq \bigr) 
- \frac{1}{7} \frac{R^5}{r^4} \Fo_a \biggr],   
\label{deltaUa}  
\end{equation} 
and we note that the terms proportional to $(R^3/r^2) \hat{\E}_{ab}$
that appear in both $\delta_1 U^{\rm out}_a$ and 
$\delta_2 U^{\rm out}_a$ cancel out after taking the sum. We note also
that the term proportional to $\Fd_a$ can be fully accounted for by
the shift in spin vector described by Eq.~(\ref{deltachi}). With the
factors of $c^{-2}$ contained in the tidal potentials, we find that 
$\delta U_a^{\rm out}$ scales as $GM/c$ and contributes to $g_{0a}$ at 
$1.5\PN$ order. 

\subsection{Light-cone coordinates} 

The post-Newtonian metric of Eq.~(\ref{PNmetric}) cannot be
compared directly with the metrics obtained in Secs.~\ref{sec:metricLC}
and \ref{sec:metricRW}, because they are expressed in different
coordinate systems. (The potentials $U$ and $U_a$ are now the full,
perturbed potentials; they were previously denoted $U + \delta U$ and  
$U_a + \delta U_a$, respectively.) Here we transform the
post-Newtonian metric to the light-cone coordinates of
Sec.~\ref{sec:metricLC}, and compare the results with the metric of 
Eq.~(\ref{metricLC}). For this purpose it is useful to look at the
post-Newtonian metric as a perturbation of the Minkowski metric, write
$g_{\alpha\beta} = \eta_{\alpha\beta} + h_{\alpha\beta}$, and work
consistently to first order in the perturbation $h_{\alpha\beta}$.  

The first step is to perform the coordinate transformation 
\begin{equation} 
ct = cv - r, \qquad x^a = r \Omega^a(\theta^A), 
\end{equation} 
which implies that $v$ is a null coordinate in flat spacetime. The
transformation brings the Minkowski metric to the form 
\begin{equation} 
\eta_{00} = -1, \qquad 
\eta_{0r} = 1, \qquad 
\eta_{AB} = r^2 \Omega_{AB}, 
\end{equation} 
and the perturbation to the form  
\begin{subequations}
\begin{align} 
h_{00} &=\frac{2}{c^2} U, \\ 
h_{0r} &= -\frac{2}{c^2} U - \frac{4}{c^3} U_a \Omega^a, \\ 
h_{0A} &= -\frac{4}{c^3} r U_a \Omega^a_A, \\
h_{rr} &= \frac{4}{c^2} U + \frac{8}{c^3} U_a \Omega^a, \\ 
h_{rA} &= \frac{4}{c^3} r U_a \Omega^a_A, \\
h_{AB} &= \frac{2}{c^2} r^2 U \Omega_{AB}. 
\end{align} 
\end{subequations} 
The second step is to perform a gauge transformation, 
\begin{equation} 
h_{\alpha\beta} \to h'_{\alpha\beta} = h_{\alpha\beta} 
- \nabla_\alpha \xi_\beta - \nabla_\beta \xi_\alpha, 
\end{equation} 
to ensure that $v$ remains null in the perturbed spacetime. This
requires $h'_{0r} = h'_{rr} = h'_{rA} = 0$, and a simple computation
reveals that the gauge vector $\xi_\alpha$ is determined by  
\begin{subequations} 
\label{LCgauge1} 
\begin{align} 
\partial_r \xi_0 &= -\frac{2}{c^2} U - \frac{4}{c^3} U_a \Omega^a, \\
\partial_r \xi_r &= \frac{2}{c^2} U + \frac{4}{c^3} U_a \Omega^a, \\
r^2 \partial_r \bigl( r^{-2} \xi_A \bigr) &= 
\frac{4}{c^3} r U_a \Omega^a_A - \partial_A \xi_r. 
\end{align} 
\end{subequations} 
With the null conditions satisfied, the remaining components of the
metric perturbation become 
\begin{subequations} 
\label{LCgauge2} 
\begin{align} 
h'_{00} &= \frac{2}{c^2} U, \\ 
h'_{0A} &= -\frac{4}{c^3} r U_a \Omega^a_A - \partial_A \xi_0, \\ 
h'_{AB} &= \frac{2}{c^2} r^2 U \Omega_{AB} - D_A \xi_B - D_B \xi_A 
- 2 r (\xi_0 + \xi_r) \Omega_{AB}. 
\end{align} 
\end{subequations} 

We focus our attention on the piece of the metric associated with
$\delta U_a$, which captures the coupling between $\chi^a$ and
$\E_{ab}$. We insert Eq.~(\ref{deltaUa}) within Eq.~(\ref{LCgauge1})
and obtain  
\begin{subequations} 
\begin{align} 
\delta \xi_0 &= \frac{75}{2} \frac{GM}{c^2} \biggl( 
\frac{1}{105} \frac{R^5}{r^3} \Ehatq + \beta_0 \biggr), \\ 
\delta \xi_r &= \frac{75}{2} \frac{GM}{c^2} \biggl( 
-\frac{1}{105} \frac{R^5}{r^3} \Ehatq + \beta_r \biggr), \\ 
\delta \xi_A &= \frac{75}{2} \frac{GM}{c^2} \biggl(  
\frac{1}{30} R^3 \Fd_A 
- \frac{1}{28} \frac{R^5}{r^2} \Fo_A 
+ r \partial_A \beta_r + r^2 \beta_A \biggr), 
\end{align}
\end{subequations} 
where $\beta_0$, $\beta_r$, and $\beta_A$ are arbitrary functions
of $\theta^A$. For our purposes here it is sufficient to restrict the
residual gauge freedom to functions of the form     
\begin{equation} 
\beta_0 = \pq_0 R^2 \Ehatq, \qquad 
\beta_r = \pq_r R^2 \Ehatq, \qquad 
\beta_A = \pd R \Fd_A + \po R \Fo_A, 
\end{equation} 
in which $\pq_0$, $\pq_r$, $\pd$, and $\po$ are arbitrary
dimensionless coefficients. Inserting the gauge vector within
Eq.~(\ref{LCgauge2}) produces 
\begin{subequations} 
\begin{align} 
\delta h'_{0A} &= \frac{75}{2} \frac{GM}{c^2} \biggl[
\frac{1}{15} \frac{R^3}{r} \Fd_A 
- 2\pq_0 R^2\, \Ehatq_A 
- \frac{1}{7} \frac{R^5}{r^3} \Fo_A \biggr], \\ 
\delta h'_{AB} &= \frac{75}{2} \frac{GM}{c^2} \biggl[
2 \bigl( 2 \pq_r - \pq_0 \bigr) R^2 r \Omega_{AB}\, \Ehatq  
- 2 \pq_r R^2 r\, \Ehatq_{AB} 
+ \biggl( \frac{1}{14} \frac{R^5}{r^2} - 2 \po R  r^2 \biggr) \Fo_{AB}
\biggr]. 
\end{align} 
\end{subequations} 
We notice that terms proportional to $(R^5/r^3) \Ehatq_A$, which used
to appear in $\delta h_{0A}$, no longer appear in $\delta h'_{0A}$. We
can further eliminate the term proportional to 
$\Omega_{AB} \Ehatq$ in $\delta h'_{AB}$ by setting $2\pq_r = \pq_0$; 
this constitutes a refinement of the light-cone gauge.     
  
By comparing $\delta h'_{0A}$ and $\delta h'_{AB}$ with the metric of 
Eq.~(\ref{metricLC}), we can read off the post-Newtonian expressions
for the relevant radial functions. We have 
\begin{subequations} 
\label{radialPN} 
\begin{align} 
\ehatq_4 &= -75 \pq_0 \frac{GM}{c^2} \frac{R^2}{r^3}, \\
\ehatq_7 &= -\frac{75}{2} \pq_0 \frac{GM}{c^2} \frac{R^2}{r^3}, \\ 
\fd_4 &= -\frac{5}{2} \frac{GM}{c^2} \frac{R^3}{r^4}, \\
\fo_4 &= -\frac{75}{14} \frac{GM}{c^2} \frac{R^5}{r^6}, \\ 
\fo_7 &= -\frac{75}{28} \frac{GM}{c^2} \biggl( 
\frac{R^5}{r^6} - 28 \po \frac{R}{r^2} \biggr). 
\end{align} 
\end{subequations} 
Comparing these with the expressions listed in Table~\ref{tab:radialLC},
we find a precise match at this leading, $1.5\PN$ order. Our
expression for $\ehatq_4$ matches the $\frac{1}{12} x^{-3} \gam{q}$ 
term in Table~\ref{tab:radialLC}, and we see that the constant $\pq_0$
can be related to $\gam{q}$. Similarly, our $\ehatq_7$ corresponds
to the $\frac{1}{24} x^{-3} \gam{q}$ term in
Table~\ref{tab:radialLC}. Our expression for $\fd_4$, which reflects
the shift of the body's angular-momentum vector created by the tidal
perturbation, matches the expected $x^{-4} \ffbar{d}$ term from
Table~\ref{tab:radialLC}, and we recall that the parameter $\ffbar{d}$
was indeed related to a shift of $\chi^a$ in Sec.~\ref{sec:metricLC}. 
Our expression for $\fo_4$ matches the expected $2 x^{-6} \ff{o}$ term
from Table~\ref{tab:radialLC}, and in this case we can assign the
precise value    
\begin{equation} 
\biggl( \frac{GM}{c^2} \biggr)^5 \ff{o} = -\frac{75}{1792} R^5 
\end{equation} 
to the associated rotational-tidal Love number. This assignment is
confirmed by comparing $\fo_7$ with the expression of
Table~\ref{tab:radialLC}, and in addition, we see that the gauge
constant $\po$ can be related to $\gam{o}$. 

Our calculation cannot produce an expression for $\ehatq_1$, which
occurs at a higher post-Newtonian order, and it also cannot produce
the terms that couple $\chi^a$ to $\B_{ab}$ in the metric of
Eq.~(\ref{metricLC}). In addition, our calculation did not produce the
expected $-2x^{-5} \ee{q}$ term in $\ehatq_4$, nor the expected 
$\frac{2}{3} x^{-5} \ee{q}$ term in $\ehatq_7$. The reason for this
must be that these terms occur at a higher post-Newtonian order. A
match at $1.5\PN$ order would have implied a scaling of the form
$(GM/c^2)^5 \ee{q} \propto (GM/c^2) R^4$ for $\ee{q}$. In the absence
of such a match, we may expect that the term appears instead at
$2.5\PN$ order, with an expected scaling of 
$(GM/c^2)^5 \ee{q} \propto (GM/c^2)^2 R^3$ for $\ee{q}$.    

As a final note, we mention that our post-Newtonian calculation did
not produce the expected $\frac{1}{2} x^{-1}\gam{d}$ term in
$\fd_4$. The fault here does not lie with a failure to go to a
sufficiently high post-Newtonian order. It lies instead with the
omission of a dipolar solution to $\nabla^2 U_a = 0$ as an additional
piece of the vector potential of Eq.~(\ref{deltaUa}). It is easy to
show that the homogeneous term 
$\delta_{\rm h} U_a = \frac{1}{4} \gam{d} GMc\, r \Fd_a$ is indeed the 
required missing piece.     

\subsection{Conclusion: Scaling of rotational-tidal Love numbers} 

The consistency between the post-Newtonian metric obtained in this
section and the exact metric obtained in Sec.~\ref{sec:metricLC} is
pleasing and reassuring, but our main purpose was not to demonstrate
this consistency. Our goal was instead to seek guidance in the scaling
of $\ee{q}$, $\bb{q}$, $\ff{o}$, $\kk{o}$,  the new class of
rotational-tidal Love numbers, with the body's radius $R$. The 
calculations presented here allow us to anticipate that for strongly
self-gravitating bodies with arbitrary internal structure, $\ff{o}$
will definitely scale as $R^5$, and $\ee{q}$ will likely scale as
$R^3$. 

Our calculations give us no direct guidance regarding $\kk{o}$
and $\bb{q}$, but inspection of the equations of post-Newtonian theory
(including the statement of hydrostatic equilibrium) indicates that at
leading order, the coupling between $\chi^a$ and $\B_{ab}$ produces
terms of order $c^{-7}$ in both $g_{00}$ and $g_{ab}$. Matching this
observation with the $x^{-6} \kk{o}$ terms in Table~\ref{tab:radialLC}
produces the expectation that $\kk{o}$ should scale as $(GM/c^2)^6
\kk{o} \propto (GM/c^2)^2 R^4$. Similarly, matching the $c^{-7}$
scaling with the $x^{-5} \bb{q}$ terms in Table~\ref{tab:radialLC}
returns an expected scaling of $(GM/c^2)^5 \bb{o} \propto (GM/c^2)^2
R^3$ for $\bb{q}$. Either one (but not both) of these estimates could
be off if the suppression of post-Newtonian order observed in the case
of $\ee{q}$ also occurred here. Thus, for example, if the $\bb{q}$ 
terms appeared only at order $c^{-9}$ in the post-Newtonian metric,
then $\bb{q}$ would scale instead as $(GM/c^2)^5 \bb{o} \propto
(GM/c^2)^3 R^2$.   

Based on this combination of definite results and educated guesswork,
we obtain the expected scalings of the rotational-tidal Love numbers
with the body's radius $R$. We express this as  
\begin{equation} 
\ee{q} =  \eee{q} \biggl( \frac{R}{GM/c^2} \biggr)^3, \qquad 
\ff{o} = \fff{o} \biggl( \frac{R}{GM/c^2} \biggr)^5, \qquad 
\bb{q} =\bbb{q} \biggl( \frac{R}{GM/c^2} \biggr)^3, \qquad 
\kk{o} = \kkk{q} \biggl( \frac{R}{GM/c^2} \biggr)^4, 
\end{equation}    
where $\eee{q}$, $\fff{o}$, $\bbb{q}$, and $\kkk{o}$ are scalefree 
versions of the rotational-tidal Love numbers. These are expected to 
be approximately independent of $GM/(c^2 R)$ when the body is weakly 
self-gravitating, but to acquire a dependence upon this quantity when
the internal gravity becomes strong. We recall the caveat made in the 
previous paragraph: our considerations cannot guarantee that the
expected scalings of $\bb{q}$ and $\kk{o}$ (and even $\ee{q}$) are not 
altered by a suppression of post-Newtonian order.    

\begin{acknowledgments} 
Useful conversations with Paolo Pani are gratefully acknowledged. One
of us (EP) is beholden to the Canadian Institute of Theoretical
Astrophysics for its warm hospitality during a research leave from the 
University of Guelph. This work was supported by the Natural Sciences
and Engineering Research Council of Canada.     
\end{acknowledgments}    

\bibliography{../bib/master}

\end{document}